\documentstyle[preprint,revtex]{aps}
{\makeatletter%
\gdef\labeleqs#1{{%
\edef\@currentlabel{%
\ifappendixon\appletter\fi
\ifsecnumbers\ifnum\c@secnum>0
\arabic{secnum}.\fi\fi\arabic{equation}}%
\label{#1}%
}}%
}
\begin{document}
\draft
\preprint{IFUP-TH 34/92}
\begin{title}
Topological susceptibility and string tension
in the lattice ${\rm CP}^{N-1}$ models
\end{title}
\author{Massimo Campostrini, Paolo Rossi, and Ettore Vicari}
\begin{instit}
Dipartimento di Fisica dell'Universit\`a and
Istituto Nazionale di Fisica Nucleare, I-56126 Pisa, Italy
\end{instit}
\begin{abstract}
In the lattice ${\rm CP}^{N-1}$ models
we studied the problems related to the measure of
observables closely connected to the dynamically generated
gauge field, such as the
topological susceptibility and the string tension.
We perfomed numerical simulations at $N=4$ and $N=10$.
In order to test the universality, we adopted two different
lattice formulations.

Scaling and universality tests led to the conclusion that
at $N=10$ the  geometrical approach gives a good definition
of lattice topological susceptibility.
On the other hand, $N=4$ proved not to be large enough to suppress
the unphysical configurations, called dislocations,
contributing to $\chi_t^g$ (at least up to
$\xi\simeq 30$ in our lattice formulations).
We obtained other determinations of $\chi_t$ by the
field theoretical method, wich relies on a local definition
of the lattice topological charge density,
and the cooling method.
They gave quite
consistent results, showing scaling and universality.

The large-$N$ expansion predicts an exponential area law behavior for
sufficiently large Wilson loops, which implies confinement, due to the
dynamical matter fields and absence of the screening phenomenon. We
determined the string tension, without finding evidence of screening
effects.

\end{abstract}


\narrowtext
\section{Introduction}
\label{Introduction}

Two-dimensional ${\rm CP}^{N-1}$ models play an important role as a
theoretical laboratory for testing non-perturbative analytical and
numerical methods in a confining, asymptotically free quantum field
theory.  A pleasant feature of these models is the possibility of
performing a systematic $1/N$ expansion around the large $N$ saddle
point solution.  Indeed, most properties of ${\rm CP}^{N-1}$ models
have been obtained in the context of the $1/N$ expansion
\cite{DiVecchia-Luscher,Witten,noiCPN-arti}.

An alternative and more general non-perturbative approach
is the simulation of the theory on the lattice.
Recently there has been considerable interest in simulations of
lattice ${\rm CP}^{N-1}$ models
\cite{CPNlatt,Wiese,Hasenbusch,Hasenbusch2,Michael,%
Michael2,Wolffcp3,Muller}.

The purpose of the present paper is that of presenting rather
complete numerical results concerning the ${\rm CP}^3$ and the
${\rm CP}^9$ models.  We especially analyze the problems related to
the measure of observables closely connected to the dynamical gauge
field, such as the topological susceptibility and the string tension.

A troublesome point in the lattice simulation technique is the study
of the topological properties.
Measuring the topological susceptibility $\chi_{\rm t}$
from discrete configurations proves to be a non-trivial task.

It is well known that in the 2-d ${\rm O}(3)$ $\sigma$ model or
${\rm CP}^1$ model, geometrical definitions of topological charge are
plagued by the presence of dislocations
\cite{Berg-Luscher,Luscher-arti}, i.e.\ topological structures of the
size of one lattice spacing, whose unphysical contribution to
$\chi_{\rm t}$ does not vanish in the continuum limit.  As a
consequence, the topological susceptibility derived from these
definitions does not show the expected scaling behavior.

The general belief is that for higher values of $N$ the above problems should
disappear. But the situation for the ${\rm CP}^3$ model appears
still problematic. There have been recent attempts
\cite{Wiese,Hasenbusch2,Michael,Wolffcp3} to determine $\chi_{\rm t}$
by using the geometrical method, but the results are not consistent.
The authors of Ref.\ \cite{Michael} and Ref.\ \cite{Wolffcp3} both
claim to observe scaling but they find different values for
$\chi_{\rm t}$ (about a factor two of difference).  Since they use
two different lattice formulations, the geometrical measure would
violate universality.  Furthermore, the authors of Ref.\
\cite{Hasenbusch2}, using the same lattice formulation as in Ref.\
\cite{Wolffcp3}, do not even see scaling.

An alternative approach relies on a definition of the topological
charge density by a local polynomial in the lattice variables.  Local
operator definitions are not affected by the dislocation problem but
unavoidably lead to mixing with lower and equal dimension operators
and to the need of subtracting perturbative tails and performing
finite renormalizations \cite{DiGiacomo-O3}.  Neverthless, this
method allowed for the determination of the topological
susceptibility of the ${\rm CP}^1$ or ${\rm O}(3)$ $\sigma$ model
\cite{CPNlatt,DiGiacomo-O3}.

Another important property of the ${\rm CP}^{N-1}$ models is the
appearance of a linear confining potential between non
gauge-invariant states.  The large-$N$ expansion predicts an
exponential area law behavior for sufficiently large Wilson loops.
The persistence of the area law at large distance would imply the
absence of the screening phenomenon due to the dynamical ``matter
fields''.  The point we wish to clarify is whether at finite $N$ the
screening phenomenon is recovered, or the large-$N$ prediction is
confirmed.

This paper is organized as follows.

In Sec.\ \ref{Lattice} the lattice actions adopted for
numerical simulations are presented and the lattice definitions
of physical observables are introduced.

In Sec.\ \ref{simulation} we discuss specific features of the
simulations of the ${\rm CP}^3$ and ${\rm CP}^9$ models, and present
the corresponding numerical results.

In Sec.\ \ref{Topology} problems related to the evaluation of the
topological susceptibility are carefully analyzed.

In Sec.\ \ref{Wilson} we discuss the determination of the string
tension from the Wilson loops.

\section{Lattice formulation}
\label{Lattice}

We choose to regularize the theory on the lattice by considering
the following action:
\begin{equation}
S_{\rm g} = -N\beta\sum_{n,\mu}\left(
   \bar z_{n+\mu}z_n\lambda_{n,\mu} +
   \bar z_nz_{n+\mu}\bar\lambda_{n,\mu} - 2\right),
\label{basic}
\end{equation}
where $z_n$ is an $N$-component complex scalar field, constrained by
the condition
\begin{mathletters}
\begin{equation}
\bar z_nz_n = 1
\end{equation}
and $\lambda_{n,\mu}$ is a ${\rm U}(1)$ gauge field satisfying
\begin{equation}
\bar\lambda_{n,\mu}\lambda_{n,\mu} = 1\,.
\end{equation}
\end{mathletters}
We also considered its tree Symanzik improved counterpart
\cite{Symanzik}
\widetext
\begin{eqnarray}
S_{\rm g}^{\rm Sym} =
-N\beta \Biggl[&&{4\over3}\sum_{n,\mu}\left(
   \bar z_{n+\mu}z_n\lambda_{n,\mu} +
   \bar z_nz_{n+\mu}\bar\lambda_{n,\mu} - 2 \right)  \nonumber \\
&&-{1\over12}\sum_{n,\mu}\left(
   \bar z_{n+2\mu}z_n\lambda_{n,\mu}\lambda_{n+\mu,\mu} +
   \bar z_nz_{n+2\mu}\bar\lambda_{n,\mu}\bar\lambda_{n+\mu,\mu} - 2
\right) \Biggr] .
\label{basicSym}
\end{eqnarray}
\narrowtext
Tests of rotation invariance and stability of adimensional ratios of
physical quantities showed that the above actions lead to scaling for
rather small correlation lengths \cite{CPNlatt}.
Comparison of measurements performed using these two actions will
provide a check of universality, implying that the two actions are
different regularizations of a unique quantum field theory.

Since the two actions are linear with respect to each lattice
variable, it is easy to construct efficient local algorithms based on
overrelaxation procedures. In our simulations we employed algorithms
consisting in efficient mixtures of over-heat bath
\cite{Petronzio-Vicari} and microcanonical \cite{Creutz} algorithms.
The detailed description of this simulation algorithm with a
discussion of its dynamical features is contained in Ref.\
\cite{CPNlatt}.

An important class of observables can be constructed by considering
the local gauge-invariant composite operator
\begin{equation}
P_{ij}(x) = \bar z_i(x) z_j(x)
\label{P-ij}
\end{equation}
and its group-invariant correlation function
\begin{equation}
G_P(x) = \left<\mathop{\rm Tr} P(x) P(0)\right>_{\rm conn} \,.
\label{G-P}
\end{equation}

The standard correlation length $\xi_{\rm w}$ is extracted from the
long-distance behavior of the zero space momentum correlation
function (``wall-wall'' correlation).
The expected large-distance behavior, including periodic boundary
condition effects, is
\widetext
\begin{equation}
G_{\rm w}(x) \simeq {A_{\rm w}\over2} \left[
   \exp\!\left(-{x\over\xi_{\rm w}}\right) +
   \exp\!\left(-{L-x\over\xi_{\rm w}}\right)\right]
\quad{\rm for}\quad {L\over2}>x\gg\xi_{\rm w} \,,
\label{corrfit-w}
\end{equation}
Moreover we measured  the ``diagonal
wall-wall'' correlation length $\xi_{\rm d}$, obtained by summing on
the correlations between points located on two distinct parallel
lines oriented at $45^\circ$ with respect to the coordinate axes,
whose large distance behavior should be
\begin{equation}
G_{\rm d}(x) \simeq {A_{\rm d}\over2} \left[
   \exp\!\left(-{x\over\xi_{\rm d}}\right) +
   \exp\!\left(-{L/\sqrt2-x\over\xi_{\rm d}}\right)\right]
\quad{\rm for}\quad {L\over2\sqrt2}>x\gg\xi_{\rm d} \,.
\label{corrfit-d}
\end{equation}
\narrowtext
In practice, $\xi_{\rm w}$, $\xi_{\rm d}$, $A_{\rm w}$, and
$A_{\rm d}$ will be obtained by fitting the data for $G_{\rm w}$ and
$G_{\rm d}$ by the functions (\ref{corrfit-w}) and (\ref{corrfit-d}),
using all the values of $x$ larger than a value $x_{\rm min}$ to be
determined.  The comparison between $\xi_{\rm w}$ and $\xi_{\rm d}$
provides a test of rotation invariance. Indeed in the scaling region,
rotation invariance implies $\xi_{\rm w} = \xi_{\rm d}$ and
$A_{\rm w}=A_{\rm d}$.  Both $\xi_{\rm w}$ and $\xi_{\rm d}$ should
reproduce in the continuum limit the inverse mass gap.

An alternative definition of the correlation length $\xi_G$ comes
from considering the second moment of the correlation function $G_P$.
In the small momentum regime we expect the behavior
\begin{equation}
\widetilde G_P(k) \approx
{Z_P \over \xi_G^{-2} + k^2}\,,
\end{equation}
where $\widetilde G_P(k)$ is the Fourier-transformed of $G_P(x)$.
The zero component of $\widetilde G_P(k)$ is by definition the
magnetic susceptibility $\chi_{\rm m}$.
On the lattice we can use the two lowest
components of $\widetilde G_P(k)$
to obtain the following definition of $\xi_G$:
\begin{equation}
\xi_G^2 = {1\over4\sin^2\pi/L} \,
\left[{\widetilde G_P(0,0)\over\widetilde G_P(0,1)} - 1\right].
\label{xiG}
\end{equation}
In the scaling region the ratio $\xi_G/\xi_{\rm w}$ must be a
constant, scale-independent number.
The large-$N$ expansion predicts \cite{noiCPN-lett}
\begin{equation}
{\xi_{\rm G}\over\xi_{\rm w}}\to \sqrt{{2\over 3}}
\label{ratioxi}
\end{equation}
when $N\to\infty$, while for $N=2$ the ratio is
equal to 1 within 1\% \cite{CPNlatt}.

The quantity $Z_P = \chi_{\rm m}\xi_G^{-2}$ is
related to the renormalization of the composite operator
$P$. Its dependence on $\beta$ can therefore be determined by
renormalization group considerations.
One finds that
\begin{equation}
Z_P = c \beta^{-2} \left[1 + O\left(1\over\beta\right)\right]\,,
\label{Z_P}
\end{equation}
where $c$ is a constant independent of the regularization scheme
and therefore of the lattice action.
In the large $N$ limit it turns out to be \begin{equation}
c = {3\over 2\pi}\;\left[1 + {8.5414\over N} +
O\left(1\over N^2\right)\right]\,.
\end{equation}

In Ref.\ \cite{CPNlatt} the quantity $A_G = Z_P\xi_{\rm w}$
was introduced. The adimensional ratio $A_{\rm w}/A_G$ is
another scheme independent quantity,
which is approximately equal to 1 in the ${\rm CP}^1$ (or
${\rm O}(3)$ $\sigma$) model \cite{CPNlatt} and
goes to zero in the large-$N$ limit because the
$\bar z z$ state becomes deconfined.

Rotation invariance and stability of adimensional physical quantities
characterize the scaling region.  Asymptotic scaling is only needed
to extract the $\Lambda$ parameter of the lattice, and to check the
predictions of the perturbation theory around the critical point.  Of
course, the requirement of scaling is weaker than that of asymptotic
scaling, which is expected to be accurately testable only much closer
to the critical point.  On the other hand, theoretically and for all
numerical experiment purposes, the scaling property is already
sufficient to simulate the physics of the continuum.

Asymptotic scaling requires the ratio of any dimensional quantity to
the appropriate power of the two-loop lattice scale
\begin{equation}
\Lambda_L \propto (2\pi\beta)^{2/N}\exp (-2\pi\beta)
\label{Lambda}
\end{equation}
to go to a constant as $\beta^{-1}\to 0$ with a linear dependence on
$\beta^{-1}$.
Furthermore, the ratio of the $\Lambda$  parameters of two
different regularizations is determined by a one loop calculation in
perturbation theory.
The ratio of the $\Lambda$ parameters of the actions (\ref{basic})
and (\ref{basicSym}) is \cite{CPNlatt}
\begin{equation}
{\Lambda_{\rm g}^{\rm Sym}\over\Lambda_{\rm g}} =
1.345\,\exp\left(0.444\over N\right)\,.
\label{ratiolambda}
\end{equation}

In the following we will also consider a modified bare coupling
extracted from the energy \cite{Parisi1,Parisi2}.
The perturbative expansion of the internal energy is
\begin{equation}
E  = {1\over 2\beta} + {2N-1\over 16N^2\beta^2}
  + O\left(1\over \beta^3\right)
\label{eneSg}
\end{equation}
for the action $S_{\rm g}$.
Using the first term we can define
\begin{equation}
\beta_E = {1\over 2E}\,.
\label{betaE}
\end{equation}
$\beta_E$ may be used as an alternative bare coupling. The first two
terms of the perturbative expansion of the $\beta$-function are
universal, therefore the asymptotic scaling function in this new
scheme is still given by Eq.\ (\ref{Lambda}).  Substituting $\beta$
with $\beta_E$ in the two loop formula (\ref{Lambda}) should be
equivalent to a resummation procedure which may improve the
asymptotic behavior.  From the Eq.\ (\ref{eneSg}) we obtain the ratio
of $\Lambda_E$, \ the $\Lambda$ parameter of the $\beta_E$ scheme,
and $\Lambda_{\rm g}$:
\begin{equation}
{\Lambda_E\over\Lambda_{\rm g}} =
\exp\left[\pi(2N-1)\over 4N^2\right]\,.
\label{ratiolambdaE}
\end{equation}
The same can be done within the action $S_{\rm g}^{\rm Sym}$.
In this case the perturbative expansion of the internal energy is
\begin{equation}
E  = {1\over 2\beta} + {0.1001N-0.0589\over N^2\beta^2} +
O({1\over \beta^3})\,,
\label{eneSgSym}
\end{equation}
and therefore we find
\begin{equation}
{\Lambda_E^{\rm Sym}\over\Lambda_{\rm g}^{\rm Sym}} =
\exp\left[\pi(0.400N-0.236)\over N^2\right]\,,\;\;\;\;\;
{\Lambda_E^{\rm Sym}\over\Lambda_{\rm g}} =
1.345\exp\left[\pi(0.844N-0.236)\over N^2\right]\,.
\label{ratiolambdaESym}
\end{equation}

\section{Simulations}
\label{simulation}

We performed Monte Carlo simulations of the ${\rm CP}^3$ model in the
two formulations (\ref{basic}) and (\ref{basicSym}) for several
values of $\beta$ corresponding to correlation lengths $\xi$ up to 30
lattice units.  A summary of the runs is presented in Table
\ref{datarun-table}.  There the integrated autocorrelation time of
the magnetic susceptibility $\tau_{\rm int}^{\chi_{\rm m}}$ is also
reported.  In the following we will also show some data for the
${\rm CP}^9$ model. Most of them were obtained from the simulations
presented in Ref.\ \cite{CPNlatt}, where the details of the runs were
given.

The data taken on lattices of different size can be used to extract
the finite size scaling functions of the correlation length
$\xi_{\rm G}$ and of the magnetic susceptibility for the ${\rm CP}^3$
model.  In the scaling region the finite size scaling functions must
be universal, that is independent of $\beta$ and of the lattice
formulation, in that they should reproduce the continuum physics in a
periodic box.  In Figs.\ \ref{FSSxi} and \ref{FSSchim} we plot
respectively $f_{\xi_{\rm G}}=\xi_{{\rm G},L}/\xi_{{\rm G},\infty}$
and $f_{\chi_{\rm m}}=\chi_{{\rm m},L}/\chi_{{\rm m},\infty}$ versus
$z=L/\xi_{\rm G}$ for the two actions (\ref{basic}) and
(\ref{basicSym}).  For both actions we chose a value of $\beta$
corresponding to a correlation length $\xi \simeq 4$.  The finite
size scaling functions were obtained by approximating infinite
lattice quantities with the corresponding values measured on the
largest lattice available.  The universality with respect to the
lattice action is fully satisfied.  From Figs.\ \ref{FSSxi} and
\ref{FSSchim} we also learn that $z\simeq 7$ is a safe value where
the finite size effects are smaller than 1\%.  The finite size
scaling functions for the ${\rm CP}^9$ model were shown in Ref.\
\cite{CPNlatt}.

Regarding the ${\rm CP}^3$ model, the data for the different
definitions of correlation length, for the correlation function
coefficient $A_{\rm w}$, and the ratio $A_{\rm w}/A_{\rm G}$ are
reported in Table \ref{xi_3-table}.  The fits to $G_{\rm w}$ and
$G_{\rm d}$ were performed choosing
$x_{\rm min}\approx 2\xi_{\rm w}$; fits using a larger $x_{\rm min}$
gave consistent results. The ratios of these different definitions
were analyzed by using the jackknife method.  The models defined by
$S_{\rm g}$ and $S_{\rm g}^{\rm Sym}$ enjoy rotation invariance and
stability of adimensional physical quantities for all values of the
correlation length considered.  The correlation lengths $\xi_{\rm G}$
and the ratios $\xi_{\rm G}/\xi_{\rm w}$ for the ${\rm CP}^9$ model
are reported in Table \ref{asysc_9-table}.  By fitting with a
constant the data of the above adimensional ratios, we obtained the
results in Table
\ref{fit-table}.

$\xi_{\rm w}$ should reproduce in the continuum limit the inverse
mass of the lowest positive parity state belonging to the adjoint
representation.  We also looked for other states, either excited
states in the adjoint positive parity channel, or states in the other
channels, the adjoint odd channel, the singlet even and odd channels.
We did not find evidence of such states for the ${\rm CP}^3$ and
the ${\rm CP}^9$ models.

Data for the constant $c$ of Eq.\ (\ref{Z_P}) are reported in Tables
\ref{asysc_3-table} and \ref{asysc_9-table} respectively for the
the ${\rm CP}^3$ and the ${\rm CP}^9$ models.  Data show scaling and
the two actions give very close values.  The small discrepancies can
be imputed to the non-universal terms of order $\beta^{-1}$ in Eq.\
(\ref{Z_P}).  We note also that the approach to scaling is slower for
the action $S_g$.

We checked the asymptotic scaling, according to the two loop formula
$f(\beta) = (2\pi\beta)^{2/N}\exp (-2\pi\beta)$, by analyzing the
quantity $M_{\rm G}/\Lambda_{\rm g}=[\xi_{\rm G}f(\beta)]^{-1}$.

To begin with, in Fig.\ \ref{asysc_1-plot} we show data for the
${\rm CP}^1$ model, which were taken by using the action $S_{\rm g}$
\cite{CPNlatt}.  If this lattice ${\rm CP}^1$ model belongs to the
universality class of the ${\rm O}(3)$ $\sigma$ model, $M_{\rm
G}/\Lambda_{\rm g}$ must tend to the asymptotic value 36.5, according
to the exact result \cite{Hasenfratz-Lambda}.
$M_{\rm G}/\Lambda_{\rm g}$ appears to be constant within the errors
for the largest values of $\beta$.  Its value (approximately 47) is
far from the asymptotic one.  However this is not a problem, since
field theory predicts an extremely slow approach to asymptopia for
quantities like $M_{\rm G}/\Lambda_{\rm g}$.  In that region of
$\beta$, corresponding to correlation lengths from about 5 to 30
lattice spacings, the $\beta$-function is well approximated by the
two loop formula (indeed $M_{\rm G}/\Lambda_{\rm g}$ is constant
within errors of $\sim 3$\%) but its integral does not (the
discrepancy is about 30\%).  It is a sort of pre-asymptotic region.

The situation becomes better if we use the $\beta_E$ scheme.
In Fig.\ \ref{asysc_1-plot} we plot also
\begin{equation}
{M_{\rm G}\over \Lambda_{\rm g}}|_E =
{M_{\rm G}\over \Lambda_E} \times {\Lambda_E\over \Lambda_{\rm g}}
\end{equation}
where the ratio $\Lambda_E/\Lambda_{\rm g}$ is obtained by using Eq.\
(\ref{ratiolambdaE}).  Now data approach the correct value,
represented in Fig.\ \ref{asysc_1-plot} by the continuous line. Is it
only an accident?

In Tables \ref{asysc_3-table} and \ref{asysc_9-table} we report data
of $M_{\rm G}/\Lambda_{\rm g}$ and $M_{\rm G}/\Lambda_{\rm g}|_E$ for
the ${\rm CP}^3$ and the ${\rm CP}^9$ model. We show them
respectively in Fig.\ \ref{asysc_3-plot} and Fig.\
\ref{asysc_9-plot}.  We use the $\Lambda$ ratios given in Eqs.\
(\ref{ratiolambda}) and (\ref{ratiolambdaE}) to report all data in
terms of $\Lambda_{\rm g}$.  Again the $\beta_E$ scheme improves the
asymptotic scaling test.  The two $\beta_E$ scheme evaluations
derived from the two actions $S_{\rm g}$ and $S_{\rm g}^{\rm Sym}$
show good agreement.  As for the ${\rm CP}^1$ case, their value is
different from those obtained with the standard schemes.

\section{Topological susceptibility}
\label{Topology}

\subsection{Introduction}

The topological charge density of a complex spin field $z$ is
\begin{equation}
q(x) = {i\over2\pi}\,\varepsilon_{\mu\nu}\, \overline{D_\mu z} D_\nu
z \,,
\end{equation}
The topological susceptibility is defined as the correlation
at zero momentum of two $q(x)$ operators:
\begin{equation}
\chi_{\rm t} = \int d^2x\,\langle q(x)\,q(0) \rangle\,.
\end{equation}
The large-$N$ predictions concerning the topological susceptibility
are \cite{Luscher-lett}
\begin{equation}
\chi_{\rm t}\xi_{\rm w}^2 = {3\over 4\pi N} +
O(N^{-5/3})\,,
\label{Luscherpred}
\end{equation}
and \cite{noiCPN-lett}
\begin{equation}
\chi_{\rm t}\xi_{\rm G}^2 = {1\over 2\pi N}\,(1-{0.38\over N}) +
O({1\over N^3})\,.
\label{chipred}
\end{equation}

Different methods have been proposed to calculate $\chi_{\rm t}$ on
the lattice. The geometrical definition uses an interpolation among
discrete lattice variables to assign an integer topological charge to
each lattice configuration.  While for large $N$ this definition is
expected to reproduce the physical topological susceptibility, at low
$N$ $\chi^{\rm g}_{\rm t}$ could receive unphysical contributions from
exceptional configurations, called dislocations, i.e.\ topological
structures of the size of one lattice spacing.  The dislocation
contributions may either survive in the continuum limit, as it
happens for some lattice formulations of the ${\rm CP}^1$ or
${\rm O}(3)$ $\sigma$ model, or push the scaling region for
$\chi^{\rm g}_{\rm t}$ to very large $\beta$ values.

Another approach relies on a definition of topological charge density
by a local polynomial in the lattice variables.
Local operator definitions are subject to mixing with lower
and equal dimension operators and to finite renormalizations,
which must be evaluated in order to extract $\chi_{\rm t}$.

A third method consists in measuring $\chi_{\rm t}$ on an ensemble of
configurations cooled by minimizing locally the action.

Any sensible definition of a lattice observable must show the
correct scaling within each lattice formulation of the theory
and universality among the determinations obtained with different
lattice actions.

\subsection{The geometrical definition}

The geometrical definition of the topological charge is
\cite{Berg-Luscher}
\FL
\begin{eqnarray}
q_n^{\rm g} = {1\over2\pi}\,\mathop{\rm Im}\bigl\{
     &&\ln[\mathop{\rm Tr} P_{n+\mu+\nu}P_{n+\mu}P_n]   \nonumber \\
  && + \ln[\mathop{\rm Tr} P_{n+\nu}P_{n+\mu+\nu}P_n]\bigr\},
\qquad \mu\ne\nu \,.
\label{geometrical-Q}
\end{eqnarray}
Introducing the quantity $\theta_{n,\mu} =
\arg\left\{\bar z_n z_{n+\mu}\right\}$, one easily obtains
\begin{equation}
q_n^{\rm g} = {1\over4\pi}\,\varepsilon_{\mu\nu} (\theta_{n,\mu} +
   \theta_{n+\mu,\nu} - \theta_{n+\nu,\mu} - \theta_{n,\nu}) \,.
\end{equation}
The periodic boundary conditions make the geometrical topological
charge of each lattice configuration, $Q_{\rm g}=\sum_n q_n^{\rm g}$,
integer.  The topological susceptibility should then be extracted by
measuring the following expectation value
\begin{equation}
\chi^{\rm g}_{\rm t}  =
{1\over V} \left< \left(Q_{\rm g}\right)^2 \right>\,.
\label{geomtop}
\end{equation}

In Table \ref{chigc_3-table} we report the data of
$\chi^{\rm g}_{\rm t}$ for the ${\rm CP}^9$ model.  Using $S_{\rm g}$
the approach to scaling is slow, instead for $S_{\rm g}^{\rm Sym}$ a
better behavior is observed.  For $S_{\rm g}$ the leading scaling
violation term must be $O(\ln \xi/\xi^2)$ when $\xi \to \infty$
\cite{Symanzik}.  Instead for the tree Symanzik improved actions the
leading logarithm corrections are absent, and scale violations are
$O(\xi^{-2})$ \cite{Paffuti}.  Assuming that the scaling violation
term proportional to $\ln\xi/\xi^2$ is already dominant in our range
of correlation lengths, we extrapolate data of $\chi_{\rm t}^{\rm g}$
for the action $S_{\rm g}$.  In Fig.\ \ref{chit_9-plot} we plot
$\chi_{\rm t}^{\rm g}$ versus $\ln \xi_{\rm G}/\xi_{\rm G}^2$. A fit
gives
\begin{equation}
\chi_{\rm t} = 0.0174(12)\,,\;\;\;b = 0.068(12)\,,
\label{chicp9}
\end{equation}
where $b$ is the coefficient of the $\ln \xi_{\rm G}/\xi_{\rm G}^2$
term.  The fitted value of $\chi_{\rm t}$ is in agreement with the
value of $\chi_{\rm t}$ obtained with the action
$S_{\rm g}^{\rm Sym}$, which is $\chi_{\rm t}=0.0176(9)$ at
$\xi_{\rm G}=5.19(3)$.  We then conclude that for the ${\rm CP}^9$
model $\chi_{\rm t}^{\rm g}$ is a good estimator of the topological
susceptibility.

The situation for the ${\rm CP}^3$ model appears more problematic.
Our data of $\chi^{\rm g}_{\rm t}$ for the ${\rm CP}^3$ model
are reported in Table \ref{chigc_3-table}
and shown in Figs.\ \ref{chit_sa-plot} and \ref{chit_ia-plot}.
For both actions an apparent scaling is observed but data
clearly violate universality.

\subsection{The field theoretical method}

The field theoretical approach
relies on a definition of topological charge density
by a local polynomial in the lattice variables having the
correct classical continuum limit
\begin{equation}
q^L(x) \to a^2 q(x) + O(a^4)
\label{classlimit}
\end{equation}
($a$ being the lattice spacing).
In order to determine $\chi_{\rm t}$, the correlation at zero
momentum of two $q^L(x)$ operators $\chi^L_{\rm t}$ is calculated:
\begin{mathletters}
\labeleqs{latticechi}
\begin{eqnarray}
\chi^L_{\rm t} &&= \biggl\langle \sum_x q^L(x)q^L(0) \biggr\rangle =
{1\over V} \left< \left(Q^L\right)^2 \right> , \\
Q^L && = \sum_x q^L(x) \,.
\end{eqnarray}
\end{mathletters}
$\chi^L_{\rm t}$ is connected to $\chi_{\rm t}$ by a nontrivial
relationship, since the presence of irrelevant operators of higher
dimension in $q^L(x)$ induces quantum corrections.  The classical
continuum limit of $q^L(x)$ must be corrected including a
renormalization constant $Z(\beta)$ \cite{Campo}.  Other
contributions originate from contact terms, i.e., from the limit
$x\to 0$ in Eq.\ (\ref{latticechi}).  These contact terms appear as
mixings with the trace of the energy-momentum tensor $S(x)$ and with
the identity operator $I$, which are the only available operators
with equal or lower dimension.  Therefore the relationship between
the lattice and the continuum topological susceptibility takes the
form
\FL
\begin{equation}
\chi_{\rm t}^L(\beta) = a^2 Z(\beta)^2\chi_{\rm t} + a^2 \,A(\beta)
   \langle S(x)\rangle_{\rm np} + P(\beta)\langle I \rangle
+ O(a^4) \,.
\label{chilchi}
\end{equation}
np denotes the nonperturbative part (i.e., the perturbative tail
must be subtracted). $Z(\beta)$, $P(\beta)$, and $A(\beta)$ are
ultraviolet effects, since they originate from the ultraviolet
cutoff-dependent modes.  They can be computed in perturbation theory
as series in $\beta^{-1}$.

The field theoretical method consists in measuring
$\chi_{\rm t}^L(\beta)$ by a standard Monte Carlo, evaluating
$Z(\beta)$, $A(\beta)$ and $P(\beta)$, and using Eq.\ (\ref{chilchi})
to extract $\chi_{\rm t}$.

The requirement (\ref{classlimit}) does not uniquely determine
the lattice operator. Different lattice versions can be found and
all of them should give the same physical result for $\chi_{\rm t}$,
instead the renormalization functions
$Z(\beta)$, $P(\beta)$, and $A(\beta)$ are lattice operator
dependent. We considered two versions of lattice topological charge
density operator.
\begin{equation}
q_1^L(x) = -{i\over 2\pi}\sum_{\mu\nu} \epsilon_{\mu\nu}
   {\rm Tr}\left[ P(x)\Delta_\mu^{(1)} P(x)
   \Delta_\nu^{(1)} P(x) \right]\,,
\label{localq1}
\end{equation}
where $\Delta^{(1)}$ is a symmetrized version of the finite
derivative:
\begin{equation}
\Delta^{(1)}_\mu P(x) = \case1/2 [P(x{+}\mu) - P(x{-}\mu)]\,.
\label{symm-deriv1}
\end{equation}
The second lattice operator  is
\begin{equation}
q_2^L(x) = -{i\over 2\pi}\sum_{\mu\nu} \epsilon_{\mu\nu}
   {\rm Tr}\left[ P(x)\Delta_\mu^{(2)} P(x)
   \Delta_\nu^{(2)} P(x) \right]\,,
\label{localq2}
\end{equation}
where $\Delta^{(2)}$ is another version of finite
derivative:
\begin{equation}
\Delta^{(2)}_\mu P(x) = \case2/3 [P(x{+}\mu) - P(x{-}\mu)]
- \case1/{12} [P(x{+}2\mu) - P(x{-}2\mu)]\,.
\label{symm-deriv2}
\end{equation}
$q_2^L(x)$ is a Symanzik tree-improved version of $q_1^L(x)$.

We applied the field theoretical method to determine the topological
susceptibility of the ${\rm CP}^3$ model.

In the following we will neglect the contribution of the mixing with
$S(x)$. This assumption is supported by a perturbative argument: the
perturbative series of $A(\beta)$ starts with a $\beta^{-3}$ term.
It will be a possible source of systematic error in our calculations.

\subsection{The heating method}

In order to estimate the renormalization functions in
Eq.\ (\ref{chilchi}) nonperturbatively, we applied the method
proposed in Ref.\ \cite{chi-letter}. We start from a configuration
$C_0$ carrying a definite topological charge $Q_0$ which is an
approximate minimum of the lattice action (in this sense we will call
it a ``classical'' configuration).  We heat it by a local updating
procedure in order to introduce short-ranged fluctuations, taking
care to leave intact the background topological structure. We
construct ensembles ${\cal C}_n^{(Q_0)}$ of many independent
configurations obtained by heating the starting configuration $C_0$
for the same number $n$ of updating steps, and average the
topological charge over the ensembles.  If $\xi \gg a$, there should
exist an intermediate range of $n$ where fluctuations of length
$l\sim a$ are thermalized at the given value of $\beta$ and reproduce
the renormalization effects, while fluctuations at the scale
$l\sim\xi$ are off equilibrium and still determined by the initial
configuration.  The average of $Q^L=\sum_x q^L(x)$
over the configurations in
this range of $n$ should be approximately equal to $Z(\beta)\,Q_0$.

We can also start from a constant configuration (with $Q_0=0$) and
construct other ensembles ${\cal C}_n^{(0)}$ of configurations. We
should find an intermediate region of $n$ where the measure of
$\chi_{\rm t}^L$ gives an estimate of the mixing $P(\beta)$ with the
identity operator which, being a short-ranged effect (due to the
fluctuations at $l\sim a$), is expected to be independent of the
physical topological background structure.

If we plot the values $Q^L$ averaged over ${\cal C}_n^{(Q_0)}$ and
the values of $\chi_{\rm t}^L$ averaged over ${\cal C}_n^{(0)}$ as
functions of $n$, we should observe plateaus in correspondence of the
above-mentioned intermediate ranges. The characteristics (starting
point and length) of the plateaus are determined by the phenomenon of
critical slowing down.  The renormalization functions are determined
by short-ranged fluctuations, which we do not expect to be critically
slowed down; therefore the starting point of the plateaus should be
independent of $\beta$. On the other hand, the end point of the
plateaus is reached when the Monte Carlo procedure changes the
long-ranged modes that determine the topological properties, and
critical slowing down should strongly affect these modes; therefore
the length of the plateaus should be $\beta$ dependent. This behavior
is essential for the existence of an intermediate range of $n$ where
the renormalization effects can be measured: indeed the success of
the present method for estimating $Z(\beta)$ and $P(\beta)$ strongly
relies on the distinction between the fluctuations at distance $l\sim
a$, contributing to the renormalizations, and those at $l\sim\xi$
determining the relevant topological properties.  The fluctuations at
$l\sim a $ are soon thermalized, whereas the topological charge
thermalization is much slower.

In order to check that heating does not change the background
topological structure of the initial configuration, after a given
number of
heating sweeps we cool the configurations (by locally minimizing the
action) and verify that the cooled configurations have topological
charge equal to $Q_0$.

We used as an updating procedure a 20-hit Metropolis algorithm
(tuned to 50\% acceptance), which gives a sufficiently mild
heating.

This method has been already applied to determine
the topological susceptibility
of the ${\rm CP}^1$ or ${\rm O}(3)$ $\sigma$ model
\cite{CPNlatt,DiGiacomo-O3}.
Consistency of the direct measures of $Z(\beta)$ and $P(\beta)$
with the corresponding perturbative computation has been shown
in Ref.\ \cite{DiGiacomo-O3} within a lattice formulation
of the ${\rm O}(3)$ $\sigma$ model.

We construct the initial configuration carrying topological charge
$Q_0=1$ (``lattice instanton'') starting from a discretization of the
continuum ${\rm SU}(2)$ instanton:
\begin{eqnarray}
z_1(x) &&= {x_1-\bar x_1-i(x_2-\bar x_2)
\over \sqrt{\rho^2+(x_1-\bar x_1)^2+(x_2-\bar x_2)^2}} \,,
 \nonumber \\
z_2(x) &&= {\rho\over \sqrt{\rho^2+(x_1-\bar x_1)^2 +
(x_2-\bar x_2)^2}} \,,  \nonumber \\
z_i(x) &&= 0,\;\;\;\;i=3,...N\,,  \nonumber \\
\lambda_\mu(x) &&= {\bar z(x{+}\mu) z(x) \over
 |\bar z(x{+}\mu) z(x)|} \,.
\label{instanton}
\end{eqnarray}
The parameter $\rho$ controls the size of the instanton, and $\bar x$
is its center, which we always place at the lattice center: $\bar x =
(L/2,L/2)$.  Starting from the configuration (\ref{instanton}), we
performed a few cooling steps in order to smooth over the
configuration at the lattice periodic boundary.  After this
procedure, we end up with a smooth configuration $C_0^{(1)}$ with
topological charge $Q^L \approx 1$. The geometrical topological
charge of this configuration is exactly equal to 1.

In Fig.\ \ref{Zeta-plot} we plot $Q_1({\cal C}_n^{(1)})/Q_{0,1}$ and
$Q_2({\cal C}_n^{(1)})/Q_{0,2}$, where $Q_i({\cal C}_n^{(1)})$
($i=1,2$) is the lattice topological charge $Q^L_i=\sum_x q^L_i(x)$
averaged over the ensemble ${\cal C}_n^{(1)}$, $Q_{0,i}$ is the
topological charge of the starting configuration measured by the
operator $Q^L_i$.  The data in Fig.\ \ref{Zeta-plot} were taken at
$\beta=1.12$ and for the tree Symanzik improved action.  We see
clearly a plateau starting from $n=7$ for both operators. For $n=12$
we also cooled the sample of configurations finding $Q^L_i\simeq
Q_{0,i}$ after a few cooling steps.  This value of $n$ is marked by a
dashed line in Fig.\ \ref{Zeta-plot}.  According to the above
considerations, the value of $Q^L_i$ at the plateau gives an estimate
of $Z_i(\beta)$.  We repeated this procedure for other values of
$\beta$, and for both actions (\ref{basic}) and (\ref{basicSym}).  We
checked also the dependence of the measure on the size of the
instanton $\rho$ (in the range of $\rho\sim\xi$), and of the value of
the topological charge of the initial configuration.  The behavior of
$Q_i({\cal C}_n^{(1)})/Q_{0,i}$ is always very similar to the case
reported in Fig.\ \ref{Zeta-plot}.  The results are presented in
Table \ref{Zeta-table}.  Configurations with topological charge two
were constructed by allocating two instantons of size $\rho$ at a
distance $d$.

We now proceed to the analysis of the ensembles ${\cal C}_n^{(0)}$ of
configurations obtained by heating the constant configuration
$C^{(0)}$, defined by $z(x)=(0,0,0,1)$ and $\lambda_\mu(x)=1$, for
several values of $\beta$.  In Figs.\  \ref{tail_sa-plot} and
\ref{tail_ia-plot} we plot the
average value of $\chi^L_{\rm t}$ as a function of the number $n$ of
heating steps for the operator $q_1^L$ and respectively for the
actions (\ref{basic}) and (\ref{basicSym}).  For every value of
$\beta$ we observe a plateau starting from $n\simeq25$; the plateau
is longer for higher values of $\beta$, as expected.  After $n_c$
heating sweeps (see Table \ref{Tail-table}) we cooled the sample of
configurations and found vanishing $Q^L_i$ in a few cooling steps.
After the plateau, $\chi_{\rm t}^L$ increases to reach the true
equilibrium value.  We identify the topological susceptibility
measured at the plateau $\chi^L_{\rm t,p}$ with the perturbative tail
at the given value of $\beta$.  Since $Z(\beta)$ and $P(\beta)$ have
their origin in the fluctuations at $l\sim a$, finite size
corrections are of the order of $L^{-2}$ and therefore negligible on
our lattice.
Results are reported in Table \ref{Tail-table}.

The values of $Z(\beta)$ and $P(\beta)$ obtained by this procedure
can be inserted in Eq.\ (\ref{chilchi}) to extract the physical value
of the topological susceptibility.
The results are summarized in Table
\ref{chit-table} and shown in Figs.\ \ref{chit_sa-plot} and
\ref{chit_ia-plot}.

All four classes of measures (2 operators $\times$ 2 actions) show
scaling within the errors. The small discrepancies among them should
be explained by the mixing with the trace of the energy-momentum
tensor $S(x)$ in Eq.\ (\ref{chilchi}).  Indeed the function
$A(\beta)$ is operator and action dependent and should vary slowly in
the relative small range of $\beta$ considered.  The fact that the
discrepancies are small give further support to the initial
assumption of neglecting the contribution of the mixing with $S(x)$
in our calculations.  From it we also get an idea of the systematic
error of our calculations.

\subsection{Heating and geometrical charge}

To clarify the origin of the failure of the geometrical method, we
followed the behavior of the geometrical topological susceptibility
$\chi_{\rm t}^{\rm g}$ during the heating procedure of the flat
configuration.  In Fig.\ \ref{tailgeom_sa-plot} we plot
$\chi^{\rm g}_{\rm t}({\cal C}_n^{(0)})$ when heating with the action
$S_{\rm g}$ at $\beta = 1.20$ and $\beta = 1.25$ (each ensembles
${\cal C}_n^{(0)}$ contains 1500 configurations).  Again after 40
heating sweeps we cooled the configurations finding vanishing
topological activity.  Therefore the signals we observe in Fig.\
\ref{tailgeom_sa-plot} are lattice artifacts, dislocation
contributions.  Notice that the value of $\chi_{\rm t}^{\rm g}$
during the heating procedure is not a negligible fraction of the
corresponding results obtained at the statistical equilibrium (see
Table \ref{chigc_3-table}).  Comparing the data up to $n=40$ for the
two values of $\beta$, we do not see evidence of critical slowing
down effects, which means that the modes responsible for the observed
signal are the short ranged (of the size of one lattice spacing),
as dislocations are supposed to be.

In Fig.\ \ref{tailgeom_ia-plot} we plot data for
$\chi^{\rm g}_{\rm t}({\cal C}_n^{(0)})$ obtained
by heating at $\beta =1.12$ with the action $S_{\rm g}^{\rm Sym}$.
The cooling check is again performed after 40 heating sweeps.
As before we observe an apparent topological activity after a few
heating sweeps, but now the signal after 40 sweeps is a smaller
fraction of the total signal measured at the equilibrium condition.
This should indicate that the action $S_{\rm g}^{\rm Sym}$
is less subject to the dislocation problems.

\subsection{Cooling method}

We performed an independent measure of $\chi_{\rm t}$ using
the cooling method \cite{Teper}, which consists in measuring
$\chi_{\rm t}$ on an ensemble of configurations cooled by locally
minimizing the action (starting from equilibrium configurations).
The idea behind the cooling method is that local changes should not
modify the topological properties of a configuration, and its
topological content can be extracted from the cooled configuration,
where the short-ranged fluctuations responsible of the
renormalization effects, have been eliminated.

The cooling algorithm consists in assigning to each lattice variable
$z_n$ ($\lambda_{n,\mu}$) a new value $z_n^\prime$
($\lambda_{n,\mu}^\prime$) (keeping all other variables fixed) that
minimizes the action.

To determine the topological charge of
the cooled configurations, we used the operator $q_2^L(x)$,
which turns out to be better than $q_1^L(x)$
in estimating the topological content of a smooth configuration.
This can be seen by comparing tha values of $Q_{0,1}$ and
$Q_{0,2}$ in Table \ref{Zeta-table}.
The topological susceptibility measured
on cooled configurations by Eq.\ (\ref{latticechi}),
$\chi_{\rm t}^{\rm cool}$, is seen to gradually rise up to an
extended plateau.
Our averages and
errors are estimated on the plateau measurements.

Table \ref{chigc_9-table} reports also data of some measurements of
$\chi_{\rm t}^{\rm cool}$ by cooling method for the ${\rm CP}^9$
model.  These measures are quite consistent, especially those
corresponding to the longest correlation lengths, with the
geometrical determinations.

For the ${\rm CP}^3$ model the results are reported in Table
\ref{chigc_3-table} and plotted in Figs.\ \ref{chit_sa-plot}
and \ref{chit_ia-plot}. $\chi_{\rm t}^{\rm cool}$ shows scaling and
the test of universality is satisfactory.
Furthermore, they are consistent with the measurements
obtained by the field theoretical method with the operator $q^L_1$.

\subsection{Conclusions}

The geometrical approach gives a good definition of lattice
topological susceptibility for $N=10$.
On the other hand, we showed that $N=4$ is not large enough
to suppress the unphysical configurations contributing to
$\chi_{\rm t}^{\rm g}$, at least for $\xi\le 30$.
The other methods, field theoretical and cooling,
give consistent measures of $\chi_{\rm t}$.
We finally quote for the ${\rm CP}^3$ model
$\chi_{\rm t}\xi_{\rm G}^2\simeq 0.06$ with
an uncertainty of 10-20\%.

\section{The string tension}
\label{Wilson}

\subsection{Wilson loops and finite size effects}

In the ${\rm CP}^{N-1}$ models
it is possible to define the (Abelian) Wilson loop
\begin{eqnarray}
W({\cal C}) = \prod_{n,\mu\in{\cal C}} \lambda_{n,\mu} \,.
\label{WilsonLoop}
\end{eqnarray}
The large-$N$ expansion predicts an
exponential area law behavior for sufficiently large Wilson
loops \cite{noiCPN-arti}:
\begin{equation}
W({\cal C}) \sim e^{-\sigma A({\cal C}) - \rho P({\cal C})}
\quad {\rm for} \quad A({\cal C}) \gg \xi^2 \,,
\end{equation}
where $\sigma$ is the Abelian string tension and $\rho$ is a
(renormalization-dependent) perimeter term.
This implies also that the dynamical matter fields
do not screen the linear potential at any distance.
The large-$N$ prediction for $\sigma$ is $\sigma\xi_{\rm G}^2=\pi/N$.

Starting from the rectangular Wilson loops, we can define the Creutz
ratios as
\begin{equation}
\chi(l,m) = \ln {W(l,m{-}1)\,W(l{-}1,m)
   \over W(l,m)\,W(l{-}1,m{-}1)} \,.
\end{equation}
The double ratio takes care of renormalization effects [constant and
perimeter terms in $\ln W(l,m)$]. It is therefore
easier to extract the string tension from $\chi(l,m)$.

In principle one can also define the Polyakov line and study the
correlation of two such lines, thus extracting the
particle-antiparticle potential. In practice the signal is so small
and noisy that one can hardly extract a physically meaningful number.

In order to extract the string tension from our simulations,
we should understand the behavior of the large Abelian Wilson
loop of a confining theory in a 2-d finite lattice with periodic
boundary conditions.

To this purpose, consider a simple 2-d
model whose lattice gauge field propagator $\Delta_\lambda$ is
\begin{equation}
\Delta_\lambda(k) = {1\over \hat{k}^2}\,,\;\;\;\;
\hat{k}^2_\mu = 4 \sin^2 {k_\mu\over 2}\,.
\label{naiveprop}
\end{equation}
For this model we find that
a rectangular $R\times T$ Wilson loop in a $L\times L$ lattice
with periodic boundary conditions has the following form:
\begin{equation}
\ln W(R,T) = -{1\over 2}\int {d^2 k\over (2\pi)^2}
{\sin^2 \left(\case1/2 k_1 R\right) \over
    \sin^2 \left(\case1/2 k_1\right)}\,
{\sin^2 \left(\case1/2 k_2 T\right) \over
    \sin^2 \left(\case1/2 k_2\right)}\,
{\hat k}^2\Delta_\lambda (k)
\label{Wilrep}
\end{equation}
where the integral must be evaluated by eliminating the zero mode.
The result of the integral is
\begin{equation}
\ln W(R,T) = {1\over2}RT\left( 1-{RT\over L^2}\right)
\label{Wilres}
\end{equation}
As expected the propagator defined in Eq.\ (\ref{naiveprop})
gives rise to a linear confining potential with a string tension
$\sigma=1/2$, but the finite size corrections are not small.
For the Creutz ratios $\eta(R)\equiv \chi(R,R)$ we obtain
(in the following we will consider only Creutz ratios
with equal arguments)
\begin{equation}
\eta(R)\equiv \chi(R,R) = {1\over 2}\,\left[ 1-\left(
{2R-1\over L}\right)^2\right]
\label{Crres}
\end{equation}

{}From this analysis we learn that, if there are not screening effects
in the ${\rm CP}^{N-1}$ models,
for a sufficiently large $R$
the behavior of the Creutz ratios $\eta(R)$
should be
\begin{equation}
\eta(R) \simeq \sigma\,\left[ 1-\left({2R-1\over L}\right)^2\right]
\label{Crratcpn}
\end{equation}
To compare data from different lattices it is convenient
to define a rescaled Creutz ratio
\begin{equation}
\eta_{\rm r}(R) = \eta(R)\,
    \left[ 1-\left({2R-1\over L}\right)^2\right]^{-1}
\simeq \sigma
\label{Crratres}
\end{equation}

A large-$N$ prediction for the behavior of the Creutz ratios
can be obtained by substituting in Eq.\ (\ref{Wilrep})
the following lattice regularized version
of the large-$N$ gauge field propagator:
\begin{equation}
\Delta_\lambda(k) = 2\pi\left( \hat{\zeta}
\ln {\hat{\zeta}+1\over\hat{\zeta}-1}-2\right)^{-1}\,,\;\;\;\;
\hat{\zeta} = \sqrt{1+{1\over \xi_{\rm w}^2\hat{k}^2}}\,.
\label{highNprop}
\end{equation}
Insertion of Eq.\ (\ref{highNprop}) in Eq.\ (\ref{Wilrep})
allows us to define the quantity $\eta^{(N)}(R)$ and, applying
Eq.\ (\ref{Crratres}), $\eta^{(N)}_{\rm r}(R)$.

\subsection{Monte Carlo results}

The gauge degrees of freedom are strongly fluctuating in the
numerical simulation; therefore large Wilson loops are hard to
measure.  The action $S_{\rm g}$ allows us to define improved
estimators for operators that are linear with respect to each
$\lambda_{n,\mu}$ variable, such as the Wilson loops.  Improved
estimators can be obtained by replacing each $\lambda_{n,\mu}$ with
its average $\lambda_{n,\mu}^{\rm imp}$ in the field of its
neighbors:
\begin{eqnarray}
\lambda_{n,\mu}^{\rm imp} &&=
{\int d \lambda_{n,\mu}  \lambda_{n,\mu} \exp\left[2\beta N
\mathop{\rm Re} \left(\bar z_{n+\mu} z_n
\lambda_{n,\mu}\right)\right] \over
\int d \lambda_{n,\mu} \exp\left[2\beta N
\mathop{\rm Re} \left(\bar z_{n+\mu} z_n
\lambda_{n,\mu}\right)\right] }
\nonumber \\
&& = {\bar z_{n+\mu} z_n \over |\bar z_{n+\mu} z_n|} \,
   {I_1(2\beta N |\bar z_{n+\mu} z_n|) \over
    I_0(2\beta N |\bar z_{n+\mu} z_n|)} \, ,
\label{imprest}
\end{eqnarray}
where $I_0$ and $I_1$ are modified Bessel functions.

Another way of reducing the noise is measuring the Wilson loops on
cooled configurations \cite{cooling}. Few cooling steps should leave
intact the longe range physical quantities, such as the string
tension, reducing the noise coming from the short ranged modes.
Cooling as other similar techniques, smearing and fuzzy operators,
provides a sequence of approximate improved estimators.  The Creutz
ratios measured on cooled configurations as function of the cooling
step are seen to reduce the errors and give origin to a plateau,
whose length depends on the size of the involved Wilson loops. Then
the cooling procedure starts to destroy the physical signal.  Our
averages and errors are estimated on the plateau measurements.

To begin with we present data for the ${\rm CP}^9$ model.
In Fig.\ \ref{CR_9_a-plot} we show the quantities
$\eta(R)\,\xi_{\rm G}^2$ and $\eta_{\rm r}(R)\,\xi_{\rm G}^2$
versus $r=R/\xi_{\rm G}$.  Data were obtained by using the action
$S_{\rm g}$ and at $\beta=0.8$ on a $60\times 60$ lattice.  The
improved estimators defined in Eq.\ (\ref{imprest}) allow good
measures up to $r\simeq 2$, for larger $r$ the signal becomes too
noisy. By using cooling we found clear signals up to $r\simeq 3$.  As
Fig.\ \ref{CR_9_a-plot} shows, the two sets of data are in perfect
agreement.  Starting from $r \simeq 2$ the rescaled Creutz ratios
show a clear plateau which is the evidence of the string tension.  We
find $\sigma\xi_{\rm G}^2=0.25(1)$, to be compared with the large-$N$
prediction $\sigma_N\xi_{\rm G}^2 = 0.314$.  This is not a surprise,
in that the quantitative agreement with the $1/N$ expansion can only
be reached at very large $N$, because of the very large coefficient
in the effective expansion parameter $6\pi/N$ that can be easily
extracted from a nonrelativistic Schr\"odinger equation analysis of
the linear confining potential
\cite{Witten,noiCPN-arti}.

$\sigma_N$ concerns the longe range predictions of the large-$N$
expansion.  We could also test the short distance predictions by
subtracting from $\eta^{(N)}_{\rm r}(R)$ the constant
$a=\sigma_N-\sigma$ and comparing the new curve
$\bar \eta^{(N)}_{\rm r}(R)$ with data. In Eq.\ (\ref{highNprop}) as
dimensional input we use the value of $\xi_{\rm G}$, which in the
large-$N$ limit is related to $\xi_{\rm w}$ by the relation
(\ref{ratioxi}).  The results of such calculations, $\bar
\eta^{(N)}(R)$ and $\bar \eta^{(N)}_{\rm r}(R)$ are shown in Fig.\
\ref{CR_9_a-plot} by the continuous lines.

In Fig.\ \ref{CR_9_b-plot} we test the universality.  Together with
the above data we plot the rescaled Creutz ratios $\eta_{\rm r}(R)$
measured by using the action $S_{\rm g}^{\rm Sym}$ at $\beta=0.75$ on
a $60\times 60$ lattice. These last data give
$\sigma\xi_{\rm G}^2=0.27(2)$, in agreement with the previous
measure.

In Fig.\ \ref{CR_3_a-plot} we show $\eta (R)$ and the corresponding
rescaled ones $\eta_{\rm r} (R)$ for the ${\rm CP}^3$ model.  Data
were taken with the action $S_{\rm g}^{\rm Sym}$ and at $\beta=0.95$
on a $60\times 60$ lattice and by using the cooling tecnique.  Again
starting from $r\simeq 2$ the rescaled Creutz ratios shows a plateau,
which gives a string tension $\sigma\xi_{\rm G}^2=0.31(2)$.

Fig.\ \ref{CR_3_b-plot} shows the rescaled Creutz ratios
coming from different simulations done with both actions
$S_{\rm g}$ and $S_{\rm g}^{\rm Sym}$, at several values of $\beta$
corresponding to correlation lengths from about 7 to 15
lattice spacings, and on different lattices.
The universality is fully satisfied.

In conclusion, we do not see  evidence of screening effects
(at least up to $3\xi$) confirming the qualitative picture coming
from the large-$N$ expansion.



\figure{Finite size scaling of the correlation length $\xi_G$ for the
${\rm CP}^3$ model.\label{FSSxi}}

\figure{Finite size scaling of the magnetic susceptibility
$\chi_{\rm m}$ for the ${\rm CP}^3$ model.\label{FSSchim}}

\figure{Asymptotic scaling test for $\xi_G$ in the
${\rm CP}^1$ model.\label{asysc_1-plot}}

\figure{Asymptotic scaling test for $\xi_G$ in the
${\rm CP}^3$ model.\label{asysc_3-plot}}

\figure{Asymptotic scaling test for $\xi_G$ in the
${\rm CP}^9$ model.\label{asysc_9-plot}}

\figure{Topological susceptibility
versus $\ln \xi_G/\xi_G^2$ in the ${\rm CP}^9$
model.\label{chit_9-plot}}

\figure{Summary of the topological susceptibility
determinations with the action $S_{\rm g}$ and for the
${\rm CP}^3$ model.\label{chit_sa-plot}}

\figure{Summary of the topological susceptibility
determinations with the action $S_{\rm g}^{\rm Sym}$ and for the
${\rm CP}^3$ model.\label{chit_ia-plot}}

\figure{Determination of the multiplicative renormalization
constants $Z^L_i$ for
the tree Symanzik improved action at $\beta=1.12$.\label{Zeta-plot}}

\figure{Determination of the perturbative tail $P$ of the topological
susceptibility constructed with the operator $q^L_1$ for the
action $S_{\rm g}$.
Solid lines show the plateau values.
\label{tail_sa-plot}}

\figure{Determination of the perturbative tail $P$ of the topological
susceptibility constructed with the operator $q^L_1$ for the
action $S_{\rm g}^{\rm Sym}$.
Solid lines show the plateau values.
\label{tail_ia-plot}}

\figure{Behavior of the geometrical topological susceptibility during
the heating procedure using $S_{\rm g}$.\label{tailgeom_sa-plot}}

\figure{Behavior of the geometrical topological susceptibility during
the heating procedure using $S_{\rm g}^{\rm
Sym}$.\label{tailgeom_ia-plot}}

\figure{Creutz ratios at $\beta=0.8$ with the action $S_g$.
${\rm CP}^9$ model. The dashed and continuous lines
are respectively $\bar \eta^{(N)}(R)$ and
$\bar \eta^{(N)}_{\rm r}(R)$.\label{CR_9_a-plot}}

\figure{Universal behavior of the rescaled Creutz ratios.
${\rm CP}^9$ model.\label{CR_9_b-plot}}

\figure{Creutz ratios at $\beta=0.95$ for the tree Symanzik
improved action.${\rm CP}^3$ model\label{CR_3_a-plot}}

\figure{Universal behavior of the rescaled Creutz ratios.
${\rm CP}^3$ model.\label{CR_3_b-plot}}


\mediumtext
\begin{table}
\caption{Summary of the simulation runs for the ${\rm CP}^3$ model.
Asterisk marks runs for the Symanzik improved action
(\ref{basicSym}).
We use the notation ``m,$\gamma$'' for a stochastic mixture
of microcanonical and over-heat bath updating with relative
weigth $\gamma$ (see Ref.\ \cite{CPNlatt}).}
\label{datarun-table}
\begin{tabular}{r@{}lrr@{ }lr@{}lr@{}lr@{}lr@{}l}
\multicolumn{2}{c}{$\beta$}&
\multicolumn{1}{r}{$L$}&
\multicolumn{2}{c}{stat}&
\multicolumn{2}{c}{$E$}&
\multicolumn{2}{c}{$\xi_G$}&
\multicolumn{2}{c}{$\chi_{\rm m}$}&
\multicolumn{2}{c}{$\tau^{\chi_{\rm m}}_{\rm int}$}\\
\tableline
 0&.95      & 12 &100k& m,4& 0&.5925(4)  & 3&.987(14)&
                     20&.30(8) &  6&.1(2)  \\
 0&.95      & 15 &100k& m,4& 0&.6005(3)  & 4&.239(16)&
                     22&.89(8) &  7&.1(2)  \\
 0&.95      & 18 &100k& m,4& 0&.6042(3)  & 4&.277(19)&
                     23&.52(11) &  7&.9(3)  \\
 0&.95      & 21 &100k& m,4& 0&.6058(2)  & 4&.272(22)&
                     23&.64(13) &  8&.4(3)  \\
 0&.95      & 24 &100k& m,4& 0&.6067(2)  & 4&.142(23)&
                     23&.01(12) &  7&.3(3)  \\
 0&.95      & 27 &100k& m,4& 0&.6070(2)  & 4&.129(25)&
                     22&.96(11) &  7&.3(3)  \\
 0&.95      & 30 &100k& m,4& 0&.6071(1)  & 4&.108(25)&
                     22&.84(11) &  7&.0(2)  \\
 0&.95      & 33 &200k& m,4& 0&.60725(8)  & 4&.094(20)&
                     22&.81(7) &  6&.6(2)  \\
 1&.05      & 60 &100k& m,4& 0&.53051(7)& 7&.59(8)&
                     62&.6(4) &  15&.8(8)  \\
 1&.15      & 120 &100k& m,4& 0&.47341(4)& 15&.1(3)&
                     189&.6(2.1) &  42&(3)  \\
 1&.20{\it a}  & 150 &100k& m,9& 0&.45018(4)& 20&.9(4)&
                     328&(4) &  64&(7)  \\
 1&.20{\it b}  & 150 &100k& m,4& 0&.45011(4)& 20&.2(4)&
                     317&(4) &  70&(7)  \\
 1&.25{\it a}  & 210 &100k& m,4& 0&.42944(3)& 27&.9(8)&
                     549&(10) &  $\approx 110$&  \\
 1&.25{\it b}  & 210 &100k& m,9& 0&.42940(3)& 28&.3(7)&
                     552&(11) &  $\approx 120$&  \\
\tableline
 0&.85 $^*$    & 12 &100k& m,4& 0&.6529(4)  & 3&.936(14)&
                     21&.54(7)  &  6&.0(2)  \\
 0&.85 $^*$    & 15 &100k& m,4& 0&.6578(3)  & 4&.212(17)&
                     24&.30(10) &  7&.5(3)  \\
 0&.85 $^*$    & 18 &100k& m,4& 0&.6613(3)  & 4&.188(20)&
                     24&.86(12) &  8&.3(3)  \\
 0&.85 $^*$    & 21 &100k& m,4& 0&.6631(2)  & 4&.160(21)&
                     24&.82(13) &  8&.2(3)  \\
 0&.85 $^*$    & 24 &100k& m,4& 0&.6638(2)  & 4&.088(22)&
                     24&.54(12) &  8&.2(3)  \\
 0&.85 $^*$    & 27 &100k& m,4& 0&.6640(2)  & 4&.025(25)&
                     24&.21(12) &  7&.6(3)  \\
 0&.85 $^*$    & 30 &100k& m,4& 0&.6639(2)  & 4&.060(27)&
                     24&.35(12) &  7&.7(3)  \\
 0&.85 $^*$    & 36 &100k& m,1& 0&.6641(2)  & 4&.044(32)&
                     24&.28(12) &  6&.9(2)  \\
 0&.95 $^*$    & 60 &100k& m,4& 0&.57665(8)  & 7&.36(8)&
                     63&.8(5) &  16&.6(9)  \\
 1&.00 $^*$    & 81 &100k& m,5& 0&.54199(6)  & 10&.14(13)&
                     106&.4(1.0) &  26&(2)  \\
 1&.05 $^*$    & 120 &60k& m,4& 0&.51178(6)  & 13&.5(3)&
                     175&(3) &  40&(4)  \\
 1&.07 $^*$    & 120 &160k& m,4& 0&.50081(4)  & 15&.4(2)&
                     215&(2) &  49&(4)  \\
 1&.12 $^*${\it a}    & 150 &160k& m,4& 0&.47555(3)  & 21&.2(4)&
                     360&(8) &  93&(9)  \\
 1&.12 $^*${\it b}    & 150 &140k& m,9& 0&.47550(4)  & 21&.0(4)&
                     361&(5) &  95&(10)  \\
\end{tabular}
\end{table}

\narrowtext
\begin{table}
\caption{Values and
ratios of different definitions of correlation length, and
correlation function coefficient for the ${\rm CP}^3$ model.}
\label{xi_3-table}
\begin{tabular}{r@{}lrr@{}lr@{}lr@{}lr@{}lr@{}l}
\multicolumn{2}{c}{$\beta$}&
{$L$}&
\multicolumn{2}{c}{$\xi_G$}&
\multicolumn{2}{c}{$\xi_{\rm w}$}&
\multicolumn{2}{c}{$\xi_G/\xi_{\rm w}$}&
\multicolumn{2}{c}{$\xi_{\rm d}/\xi_{\rm w}$}&
\multicolumn{2}{c}{$A_{\rm w}/A_G$}\\
\tableline
0&.95 & 33 & 4&.094(20) & 4&.151(21)&0&.985(4)&0&.999(5)&0&.950(13)\\
1&.05 & 60 & 7&.59(8) & 7&.66(14)& 0&.991(10)& 1&.014(15)& 0&.97(4)\\
1&.15 & 120 & 15&.1(3) & 15&.4(4)& 0&.980(16)& 0&.980(20)& 0&.93(5)\\
1&.20 & 150 & 20&.5(3) & 20&.7(5)& 0&.992(13)& 0&.992(17)& 0&.98(4)\\
1&.25 & 210 & 28&.1(5) & 28&.9(1.2)& 0&.973(21)& 0&.98(5)& 0&.91(7)\\
\tableline
0&.85 $^*$& 30 & 4&.06(3) & 4&.14(4)& 0&.982(6)& 0&.998(8)& 0&.936(19)\\
0&.85 $^*$& 36 & 4&.04(3) & 4&.12(3)& 0&.982(6)& 0&.984(10)&0&.935(21)\\
0&.95 $^*$& 60 & 7&.36(9) & 7&.46(10)& 0&.987(11)& 0&.991(14)&0&.95(3)\\
1&.00 $^*$& 81 & 10&.14(14) & 10&.30(20)& 0&.984(12)&0&.98(2)&0&.95(4)\\
1&.05 $^*$& 120 & 13&.5(4) & 13&.6(5)& 0&.992(21)& 0&.98(3)& 0&.97(7)\\
1&.12 $^*$& 150 & 21&.1(3) & 21&.4(6)& 0&.983(12)& 1&.01(2)& 0&.95(4)\\
\end{tabular}
\end{table}

\begin{table}
\caption{Some results for the ${\rm CP}^3$ model.}
\label{asysc_3-table}
\begin{tabular}{r@{}lr@{}lr@{}lr@{}l}
\multicolumn{2}{c}{$\beta$}&
\multicolumn{2}{c}{$\beta^2Z_P$}&
\multicolumn{2}{c}{$M_G/\Lambda_{\rm g}$}&
\multicolumn{2}{c}{$M_G/\Lambda_{\rm g}|_{E}$}\\
\tableline
 0&.95     & 1&.228(7) & 39&.1(2) & 26&.7(1)  \\
 1&.05     & 1&.20(2) & 37&.6(4) & 28&.5(3)  \\
 1&.15     & 1&.10(3) & 33&.9(6) & 27&.6(5)  \\
 1&.20     & 1&.10(2) & 33&.4(4) & 27&.9(4)  \\
 1&.25     & 1&.08(3) & 32&.7(6) & 27&.9(5)  \\
\tableline
 0&.85$^*$ & 1&.067(9) & 33&.4(2) & 25&.3(2)  \\
 0&.95$^*$ & 1&.06(2) & 32&.7(4) & 26&.6(3)  \\
 1&.00$^*$ & 1&.04(2) & 31&.7(4) & 26&.5(3)  \\
 1&.05$^*$ & 1&.05(3) & 31&.7(8) & 27&.2(6)  \\
 1&.07$^*$ & 1&.04(2) & 31&.3(4) & 27&.0(4)  \\
 1&.12$^*$ & 1&.03(2) & 30&.6(4) & 26&.9(3)  \\
\end{tabular}
\end{table}

\begin{table}
\caption{Some results for the ${\rm CP}^9$ model.}
\label{asysc_9-table}
\begin{tabular}{r@{}lr@{}lr@{}lr@{}lr@{}lr@{}l}
\multicolumn{2}{c}{$\beta$}&
\multicolumn{2}{c}{$\xi_G$}&
\multicolumn{2}{c}{$\xi_G/\xi_{\rm w}$}&
\multicolumn{2}{c}{$\beta^2Z_P$}&
\multicolumn{2}{c}{$M_G/\Lambda_{\rm g}$}&
\multicolumn{2}{c}{$M_G/\Lambda_{\rm g}|_{E}$}\\
\tableline
 0&.7   & 2&.35(3) & 0&.971(6) & 0&.905(20) & 25&.73(33) & 20&.57(27)\\
 0&.75  & 3&.31(4) &  &---     & 0&.860(20) & 24&.63(31) & 20&.46(26)\\
 0&.8   & 4&.67(3) & 0&.956(6) & 0&.830(8)  & 23&.65(14) & 20&.30(13)\\
 0&.85  & 6&.44(6) & 0&.962(10)& 0&.823(12) & 23&.18(22) & 20&.34(19)\\
 0&.9   & 8&.83(8) & 0&.973(9) & 0&.816(11) & 22&.88(21) & 20&.42(19)\\
\tableline
 0&.7$^*$  & 3&.78(2) & 0&.968(6) & 0&.795(7) & 22&.53(11)&19&.90(9) \\
 0&.75$^*$ & 5&.19(3) & 0&.974(6) & 0&.787(7) & 22&.18(13)&19&.94(12)\\
\end{tabular}
\end{table}

\begin{table}
\caption{Tests of scaling by fitting adimensional ratio data
with a constant.}
\label{fit-table}
\begin{tabular}{ccr@{}lr@{}l}
$N$ & $S$ &
\multicolumn{2}{c}{$\xi_G/\xi_{\rm w}$}&
\multicolumn{2}{c}{$A_{\rm w}/A_G$}\\
\tableline
 4    & $S_{\rm g}$           &  0&.986(3) & 0&.952(11) \\
 4    & $S_{\rm g}^{\rm Sym}$ &  0&.984(4) & 0&.944(14) \\
 10   & $S_{\rm g}$           &  0&.965(4) & 0&.870(13) \\
 10   & $S_{\rm g}^{\rm Sym}$ &  0&.971(4) & 0&.900(14) \\
\end{tabular}
\end{table}

\narrowtext
\begin{table}
\caption{Geometric and cooled topological
susceptibility for the ${\rm CP}^3$ model.}
\label{chigc_3-table}
\begin{tabular}{r@{}lrr@{}lr@{}lr@{}lr@{}l}
\multicolumn{2}{c}{$\beta$}& $L$ &
\multicolumn{2}{c}{$10^4\chi_{\rm t}^{\rm g}$}&
\multicolumn{2}{c}{$\chi_{\rm t}^{\rm g}\xi^2_G$}&
\multicolumn{2}{c}{$10^4\chi_{\rm t}^{\rm cool}$}&
\multicolumn{2}{c}{$\chi_{\rm t}^{\rm cool}\xi^2_G$}\\
\tableline
 0&.95    & 33 & 49&.5(5)   & 0&.083(1)   &   &---     &   &---     \\
 1&.05    & 60 & 16&.2(3)   & 0&.093(3)   &  8&.6(3)   &  0&.050(2) \\
 1&.15    & 120&  4&.01(11) & 0&.091(6)   &  2&.70(11) &  0&.062(3) \\
 1&.20    & 150&  2&.11(5)  & 0&.089(3)   &  1&.44(5)  &  0&.061(3) \\
 1&.25    & 210&  1&.06(3)  & 0&.084(4)   &  0&.77(4)  &  0&.062(4) \\
\tableline
 0&.85$^*$& 36 & 40&.5(6)   & 0&.066(1)   &   &---     &   &---     \\
 0&.95$^*$& 60 & 13&.4(3)   & 0&.072(2)   &  9&.1(3)   &  0&.049(2) \\
 1&.00$^*$& 81 &  7&.0(2)   & 0&.072(3)   &  5&.2(2)   &  0&.054(2) \\
 1&.05$^*$&120 &  3&.7(2)   & 0&.068(4)   &  3&.2(2)   &  0&.058(5) \\
 1&.07$^*$&120 &  2&.83(7)  & 0&.067(2)   &  2&.37(8)  &  0&.056(3) \\
 1&.12$^*$&150 &  1&.50(4)  & 0&.067(2)   &  1&.27(4)  &  0&.057(2) \\
\end{tabular}
\end{table}

\narrowtext
\begin{table}
\caption{Geometric and cooled topological
susceptibility for the ${\rm CP}^9$ model.}
\label{chigc_9-table}
\begin{tabular}{r@{}lrr@{}lr@{}lr@{}lr@{}l}
\multicolumn{2}{c}{$\beta$}& $L$ &
\multicolumn{2}{c}{$10^4\chi_{\rm t}^{\rm g}$}&
\multicolumn{2}{c}{$\chi_{\rm t}^{\rm g}\xi^2_G$}&
\multicolumn{2}{c}{$10^4\chi_{\rm t}^{\rm cool}$}&
\multicolumn{2}{c}{$\chi_{\rm t}^{\rm cool}\xi^2_G$}\\
\tableline
 0&.7     & 42 & 50&.5(1.1) & 0&.0279(9)    &   &---   &  &---      \\
 0&.75    & 60 & 22&.6(6)   & 0&.0248(9)    & 17&.8(8) & 0&.0195(10)\\
 0&.8     & 60 &  9&.9(3)   & 0&.0216(6)    &  9&.1(5) & 0&.0197(10)\\
 0&.85    & 72 &  5&.1(4)   & 0&.0213(17)   &   &---   &  &---      \\
 0&.9     & 90 &  2&.5(3)   & 0&.0198(24)   &   &---   &  &---      \\
\tableline
 0&.7 $^*$& 42 & 12&.3(6)   & 0&.0179(9)   &    &---   &  &---      \\
 0&.7 $^*$& 60 & 13&.2(5)   & 0&.0186(7)   &  12&.3(4) & 0&.0173(7) \\
 0&.75$^*$& 81 &  6&.5(3)   & 0&.0176(9)   &   6&.1(5) & 0&.0166(13)\\
\end{tabular}
\end{table}

\begin{table}
\caption{Measure of the multiplicative renormalization of $\chi^L$,
starting from an instanton of size $\rho$ or two instantons
of size $\rho$ and distance $d$.
$Q_G$ is the geometrical charge of the initial configuration.
The estimate of $Z^L_i$ are taken by averaging the data in the range of
$n$ reported in the column ``plateau''.}
\label{Zeta-table}
\begin{tabular}{r@{}lccrrrcclcl}
\multicolumn{2}{c}{$\beta$}& $L$ & $Q_G$ & $\rho$ & $d$ &
 Stat & plateau & $Q_{1,0}$ & $Z^L_1$ & $Q_{2,0}$ & $Z^L_2$ \\
\tableline
1&.15 &48&1& 10 &---& 1000 & 8--10 & 0.9908 & 0.375(7)&0.9997&0.420(12)\\
1&.15 &60&2&  8 & 12&  400 & 8--10 & 1.9165 & 0.375(8)&1.9912&0.417(11)\\
1&.15 &60&2&  8 & 16&  400 & 8--10 & 1.9517 & 0.377(7)&1.9971&0.418(10)\\
1&.20 &48&1& 10 &---& 2000 & 7--10 & 0.9908 & 0.413(5)&0.9997&0.461(7) \\
1&.20 &60&2&  8 & 20&  400 & 7--10 & 1.9590 & 0.413(7)&1.9983&0.459(10)\\
1&.20 &48&1& 10 &---& 1000 & 7--10 & 0.9908 & 0.432(6)&0.9997&0.481(10)\\
\tableline
1&.07$^*$ & 48 & 1 & 10 &---& 1000 & 7--10 & 0.9905 & 0.442(6) &
0.9997 &0.494(9) \\
1&.12$^*$ & 48 & 1 & 10 &---& 1000 & 7--12 & 0.9905 & 0.464(5) &
0.9997 &0.518(8) \\
1&.17$^*$ & 48 & 1 & 10 &---&  500 & 7--12 & 0.9905 & 0.487(7) &
0.9997 &0.541(11) \\
\end{tabular}
\end{table}

\begin{table}
\caption{Measure of the perturbative tail $P_i$ of $\chi^L_i$.
Data were taken on $36\times 36$ lattice.}
\label{Tail-table}
\begin{tabular}{r@{}lrccr@{}lr@{}l}
\multicolumn{2}{c}{$\beta$}
& Stat & plateau & $n_c$ &
\multicolumn{2}{c}{$10^4 P_1$}
&\multicolumn{2}{c}{$10^4 P_2$}
\\
\tableline
1&.15 & 5000 & 23--30 & 28 & 0&.293(6) & 0&.714(14) \\
1&.20 & 6000 & 24--40 & 35 & 0&.244(4) & 0&.592(12) \\
1&.25 & 6500 & 28--40 & 40 & 0&.202(3) & 0&.491(9) \\
1&.30 & 1000 & 22--40 & 40 & 0&.170(7) & 0&.42(2)  \\
\tableline
1&.07 $^*$ & 3000 & 24--33 & 28 & 0&.215(6) & 0&.508(13) \\
1&.12 $^*$ & 3000 & 25--38 & 40 & 0&.176(6) & 0&.414(10) \\
1&.17 $^*$ & 1000 & 22--40 & 40 & 0&.140(6) & 0&.33(2) \\
\end{tabular}
\end{table}

\mediumtext
\begin{table}
\caption{Measure of $\chi_{\rm t}$ by the field theoretical method.}
\label{chit-table}
\begin{tabular}{r@{}lr@{}lr@{}lr@{}lr@{}lr@{}lr@{}lr@{}l}
\multicolumn{2}{c}{$\beta$}&
\multicolumn{2}{c}{$10^4\chi^L_1$}&
\multicolumn{2}{c}{$10^4\chi_{\rm t}^{\rm ft,1}$}&
\multicolumn{2}{c}{$\chi_{\rm t}^{\rm ft,1}\xi_G^2$}&
\multicolumn{2}{c}{$10^4\chi^L_2$}&
\multicolumn{2}{c}{$10^4\chi_{\rm t}^{\rm ft,2}$}&
\multicolumn{2}{c}{$\chi_{\rm t}^{\rm ft,2}\xi_G^2$}& \\

1&.15  & 0&.647(18) & 2&.52(17)  & 0&.058(4) &
         1&.265(36) & 3&.1(3)    & 0&.071(7) \\
1&.20  & 0&.485(12) & 1&.41(8)   & 0&.060(4) &
         0&.952(22) & 1&.68(14)  & 0&.071(5) \\
1&.25  & 0&.345(8)  & 0&.76(5)   & 0&.060(5) &
         0&.706(15) & 0&.92(9)   & 0&.072(8) \\
\tableline
1&.07$^*$  & 0&.646(16) & 2&.21(11)  & 0&.052(4) &
             1&.147(28) & 2&.62(16)  & 0&.062(4) \\
1&.12$^*$  & 0&.442(10) & 1&.24(6)   & 0&.055(3) &
             0&.800(16) & 1&.44(9)   & 0&.064(4) \\
\end{tabular}
\end{table}

\end{document}